\documentclass[prl,twocolumn,showpacs,superscriptaddress]{revtex4}

\usepackage{anysize}
\marginsize{1.5cm}{1.5cm}{1.cm}{1.cm}
\usepackage{graphicx}
\usepackage{epsfig}
\usepackage{verbatim}
\usepackage{amssymb}
\usepackage{amsfonts,amsmath}
\usepackage{bm}
\usepackage[T1]{fontenc}
\usepackage{eufrak}
\usepackage[colorlinks]{hyperref}
\usepackage{fancyhdr}
 \usepackage{relsize}

\hypersetup{colorlinks=true,
	linkcolor=black,
	anchorcolor=black,
	citecolor=black,
	urlcolor=black
}

\providecommand{\be}{\begin{equation}}
\providecommand{\ee}{\end{equation}}
\providecommand{\bea}{\begin{eqnarray}}
\providecommand{\eea}{\end{eqnarray}}
\providecommand{\beas}{\begin{eqnarray*}}
\providecommand{\eeas}{\end{eqnarray*}}

\providecommand{\beni}{\begin{equation*}}
\providecommand{\eeni}{\end{equation*}}

\providecommand{\bw}{\begin{widetext}}
\providecommand{\ew}{\end{widetext}}

\newlength{\bilderlength}

\newlength{\figsize}
\begin{document}

\title{Extreme-Value Distributions and the Freezing Transition of Structural Glasses}

\author{Michele Castellana}
\email{michelec@princeton.edu}
\affiliation{Joseph Henry Laboratories of Physics and Lewis--Sigler Institute for Integrative Genomics, Princeton University, Princeton, New Jersey 08544}


\begin{abstract} 
  We consider two mean-field models of structural glasses, the random energy model (REM) and the $p$-spin model (PSM), and we show that the finite-size fluctuations of the freezing temperature are described by extreme-value statistics (EVS) distributions, establishing an unprecedented connection between EVS and the freezing transition of structural glasses. For the REM, the freezing-temperature fluctuations are described by the Gumbel EVS distribution, while for the PSM the freezing temperature fluctuates according to the Tracy-Widom (TW) EVS distribution, which has been recently discovered within the theory of random matrices.  For the PSM, we provide an analytical argument showing that the emergence of the TW distribution can be understood in terms of the statistics of glassy metastable states.  
\end{abstract}

\maketitle

The problem of characterizing the maximum of a set of random variables, also known as extreme-value statistics (EVS), has attracted considerable interest for several decades now. EVS has found a remarkable number of applications in several scientific fields, including engineering \cite{gumbel1958statistics}, finance \cite{embrechts1997modelling}, and biology \cite{omalley2009ecological}. Moreover, EVS has raised particular interest in physics \cite{bouchaud1997universality}, describing the phenomenology of several  systems of general interest, such as dynamical \cite{collet2001statistics}, and disordered \cite{ledoussal2003exact} systems. In particular, EVS has been recently shown to play a role in the critical regime of spin glasses, i.e. disordered uniaxial magnetic materials like $\textrm{Fe}_{0.5} \textrm{Mn}_{0.5} \textrm{TiO}_3$ and $\textrm{Eu}_{0.5} \textrm{Ba}_{0.5} \textrm{MnO}_3$ \cite{castellana2011role,castellana2011extreme}: In this Letter, we establish a novel connection between EVS and a completely different class of solid materials. These systems, known as structural glasses, are liquids--like o-Terphenyl and Glycerol--that have been cooled fast enough to avoid cristallisation \cite{tarjus2001viscous}.  

Since its very first development, the theory of structural glasses has continuously drawn the attention of the scientific community: Understanding the low-temperature behavior of these systems and the nature of their glassy phase is still one of the deepest unsolved questions in condensed-matter theory \cite{anderson1995through}. In particular, the existence of a freezing transition in structural glasses has been the subject of an ongoing debate for the last few decades \cite{biroli2009random}. The development of exactly solvable models \cite{derrida1980randomlimit,gross1984simplest} mimicking the phenomenology of structural  glasses showed that such a transition does exist on a mean-field level and, later on, further studies suggested \cite{kirkpatrick1989scaling,biroli2009random} that a freezing transition might occur also for realistic, non-mean-field \cite{castellana2010hierarchical} systems. 

We consider two well-established \cite{biroli2009random,castellani2005spin} mean-field models of the freezing transition of structural glasses, the random energy model (REM)  \cite{derrida1980randomlimit}  and the $p$-spin model (PSM) \cite{gross1984simplest}, and we study the disorder-induced fluctuations of the critical temperature arising when the system size is large but finite. We show that for the REM the  fluctuations of the critical temperature are described by a EVS distribution of independent variables: The Gumbel distribution \cite{gumbel1958statistics}. For the PSM, the finite-size fluctuations of the critical point are described by a EVS distribution of correlated variables, the Tracy-Widom (TW) distribution \cite{tracy2002proceedings}. The TW distribution has been recently discovered in the theory of random matrices, and it describes random fluctuations in a variety of physical systems \cite{somoza2007universal,takeuchi2010universal}.
\textit{Random energy model} - Let us start by considering the simplest model of a structural glass exhibiting a freezing transition: The random energy model (REM)  \cite{derrida1980randomlimit}. Here, the REM will serve as an illustrative model to show the role played by EVS in the structural-glass freezing transition.  The REM  is defined as a system of $N$ Ising spins $S_i = \pm 1$: An energy $E[\vec{S}]$ is assigned to every spin configuration $\vec{S}$, and the energies $\mathcal{E} \equiv \{E[\vec S ]\}$ are  independent and identically distributed (IID) Gaussian random variables with zero mean and variance $N/2$. In the thermodynamic limit, the REM has a freezing phase transition:  There is a critical value of the energy $e_c$, which is is the lowest value of  $e$ such that the number of states $\vec{S}$ with energy $E[\vec{S}] = N e$ is exponentially large in the system size. The inverse critical temperature $\beta_c$ is determined from the threshold energy $e_c$ by the temperature-energy relation $\beta_c = - 2 e_c$, which can be obtained by computing the Legendre transform of the partition function \cite{derrida1980randomlimit}. 
Let us now introduce a critical temperature for a REM with a finite number of spins and let us study its sample-to-sample fluctuations. If the system size $N$ is sufficiently large, the threshold energy $e_{c \, \mathcal{E}}$ of an energy sample $\mathcal{E}$  coincides with the lowest energy value, i.e.  $N e_{c \, \mathcal{E}} = \min_{\vec{S}} E[\vec{S}]$. It follows \cite{dedominicis2006random} that for large $N$ the threshold energy is $e_{c \, \mathcal{E}} = e_c- \mathlarger{\chi  / (2 N \sqrt{\log 2})}$, 
where $\chi$ is a random variable distributed according to the Gumbel distribution: $\textrm{P}(\chi \leq x) = \exp(-\exp(-x))$. It is easy to show that the Gumbel distribution describes not only the statistics of the ground state \cite{dedominicis2006random}, but also the fluctuations of the critical temperature. A natural way to introduce a finite-size critical temperature is to extend the temperature-energy relation $\beta_c = - 2 e_c$ to systems with finite sizes \cite{castellana2011extreme,castellana2011role,billoire2011finite}: We set $\beta_{c \, \mathcal{E} } = -2e_{c \, \mathcal{E}}$, where   $\beta_{c \, \mathcal{E} }$ is the finite-size critical temperature of sample $\mathcal{E}$. Putting this definition together with the above expression for $e_{c \, \mathcal{E}}$, we obtain the expression for the finite-size critical temperature
$\beta_{c \, \mathcal{E}} =  \beta_c + \mathlarger{\chi  /( N \sqrt{\log 2})}$.  This shows that the finite-size fluctuations of the critical temperature of the REM are described by a EVS distribution, the Gumbel distribution, and that the width of the critical region $\beta_{c \, \mathcal{E}} - \beta_c$ scales with the system size as $1/N$. 
Given that the REM is a highly simplified model for a structural glass \cite{biroli2009random,castellana2010hierarchical}, a natural question to ask is whether EVS could play a role in other structural-glass models more realistic than the REM. We will address this question in the following Section.

\begin{figure}
\vspace{5mm}
\includegraphics[scale=0.9]{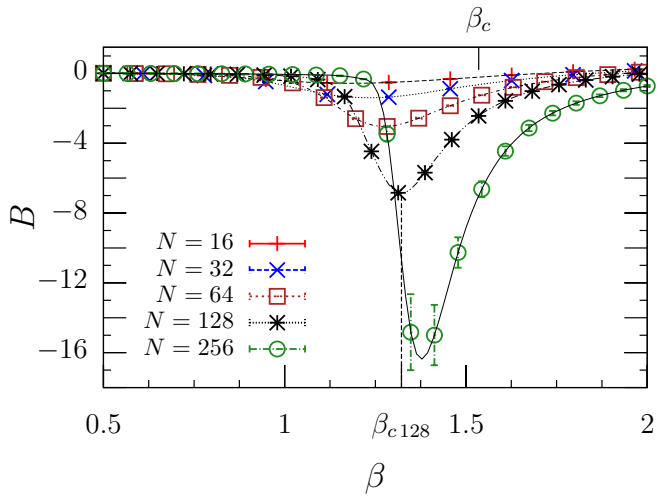}
\caption{
Binder cumulant $B$ of the $p$-spin model with $p=3$ as a function of the inverse temperature $\beta$ for system sizes $ N = 16, 32, 64, 128,  256$ (in red, blue, brown, black, green respectively),  critical temperature $\beta_{c \, N}$ for $N=128$, and infinite-size critical temperature $\beta_c$.  
\label{fig1}}
\end{figure}

\textit{$p$-spin Model} - In this Section we will consider the mean-field $p$-spin model (PSM) \cite{gross1984simplest,derrida1980randomlimit}.  The PSM captures some fundamental physical features of structural glasses that are absent in the REM  \cite{biroli2009random}: For example, in the PSM energy minima are no longer single configurations $\vec{S}$ like in the REM, but they also allow for small vibrations around the minimum \cite{biroli2009random}. Moreover, the PSM reproduces the onset of a dynamical slowing down at low temperatures typical of structural glasses \cite{castellani2005spin}. In what follows,  we will focus on the PSM with $p=3$ \cite{billoire2005mean}, which  is given by $N$ Ising spins $S_i = \pm 1$ with Hamiltonian 
$
H[\vec{S}] =  - \frac{\sqrt{3}}{N} \sum_{i<j<k} J_{ijk} S_i S_j S_k,
$
where $J_{ijk}$ are IID random variables equal to $\pm 1$ with equal probability. Unlike the REM,  the problem of studying  sample-to-sample fluctuations of the critical temperature of the PSM cannot be addressed analytically. Hence, we have studied the PSM numerically by means of Monte Carlo simulations combined with the parallel-tempering algorithm \cite{swendsen1986replica}. Even though the CPU time for the PSM increases with the system size as $N^3$ \cite{billoire2005mean}, we exploited the binary form of the couplings to use  an efficient asynchronous multispin coding method \cite{palassini1999universal}, allowing us to study extensive system sizes and numbers of disorder samples.  Specifically, we studied systems with $N = 16,32,64,128,256$ spins and a number of disorder samples $2.6 \times 10^2 \leq S \leq 1.3\times 10^5 $.

\begin{figure}
\vspace{5mm}
\includegraphics[scale=0.92]{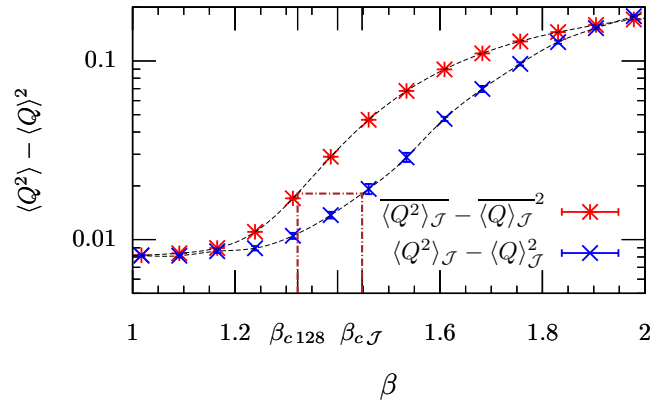}
\caption{
Average order-parameter fluctuations $ \overline{\langle Q^2 \rangle _{\mathcal{J}}}  - \overline{\langle Q \rangle _{\mathcal{J}}}^2$ (in red), and single-sample order-parameter fluctuations  $ {\langle Q^2 \rangle _{\mathcal{J}}}  - \langle Q \rangle _{\mathcal{J}}^2$ (in blue) as functions of the inverse temperature $\beta$ for the $p$-spin model with $p=3$ and $N = 128$. The value of $\beta_{c \, 128}$ is taken from Fig. \ref{fig1}, and the  inverse critical temperature $\beta_{c \, \mathcal{J}}$ is determined by Eq. (\ref{eq5}) (in brown).
\label{fig2}}
\end{figure}

 In the thermodynamic limit, the PSM is known \cite{biroli2009random} to have a mixed first/second order phase transition: Given two independent replicas $\vec{S}^1$, $\vec{S}^2$, this phase transition can be described in terms of  their mutual overlap $Q \equiv 1/N \sum_{i=1}^{N} S^1_i S^2_i$.  In the high-temperature phase $\beta <\beta_c =   1.535$ \cite{billoire2005mean}, the two replicas explore exponentially many energy minima. Since these minima are random states not related by any symmetry, different minima have  zero mutual overlap \cite{biroli2009random, castellani2005spin}, and one has $\overline{\langle Q^2 \rangle} = \overline{\langle Q \rangle} = 0$,  where $\langle \rangle$ is the Boltzmann average, and $\overline{  \phantom{X\,} }$ denotes the average over disorder samples $\{ J_{ijk} \} \equiv \mathcal{J}$.  In the low-temperature phase $\beta > \beta_c$, the two replicas are both trapped in a few low-lying energy minima, and thus they develop a nonzero mutual overlap: $\overline{\langle Q \rangle} >0$, $\overline{\langle Q^2 \rangle} - \overline{\langle Q \rangle}^2 > 0$. Let us now introduce a finite-size critical temperature for the PSM. In the first place, the above discussion of the phase transition of the PSM shows that the infinite-volume critical point $\beta_c$  is the value of the temperature at which the order-parameter fluctuations (OPF)  arise: $\overline{\langle Q^2 \rangle}  - \overline{\langle Q \rangle}^2= 0 \text{ if } \beta \lesssim \beta_c $, $\overline{\langle Q^2 \rangle}  - \overline{\langle Q \rangle}^2 > 0 \text{ if } \beta \gtrsim \beta_c$.  To introduce a critical temperature $\beta_{c \, \mathcal{J}}$ of a finite-size PSM with couplings $\mathcal{J}$, we recall that the Binder cumulant $B \equiv \frac{1}{2} (3 - \overline{\langle  Q^4\rangle} / \overline{\langle  Q^2\rangle}^2)$ of a finite-size PSM has a minimum at a given temperature: This temperature, which we will denote by $\beta_{c \, N}$, is the temperature at which critical OPF arise \cite{billoire2005mean}, and it is shown in Fig. \ref{fig1}. It follows that the quantity $\overline{\langle Q^2 \rangle(\beta_{c \, N})}  - \overline{\langle Q \rangle(\beta_{c \, N})}^2$ represents the average OPF at the critical point. Given these average critical OPF, it is natural to identify the finite-size critical temperature $\beta_{c \, \mathcal{J}}$ of sample $\mathcal{J}$ as the value of $\beta$ for which the OPF of the sample $\langle Q^2 \rangle_{\mathcal{J}}(\beta) - \langle Q \rangle_{\mathcal{J}}(\beta)^2$ are equal to the above average critical value
 \bea\label{eq5}
 \langle Q^2 \rangle_{\mathcal{J}}(\beta_{c \, \mathcal{J}}) - \langle Q \rangle_{\mathcal{J}}(\beta_{c \, \mathcal{J}})^2= \overline{\langle Q^2 \rangle(\beta_{c \, N})}  - \overline{\langle Q \rangle(\beta_{c \, N})}^2. \; \, \, &&
\eea

\begin{figure}
\vspace{5mm}
\includegraphics[scale=1.25]{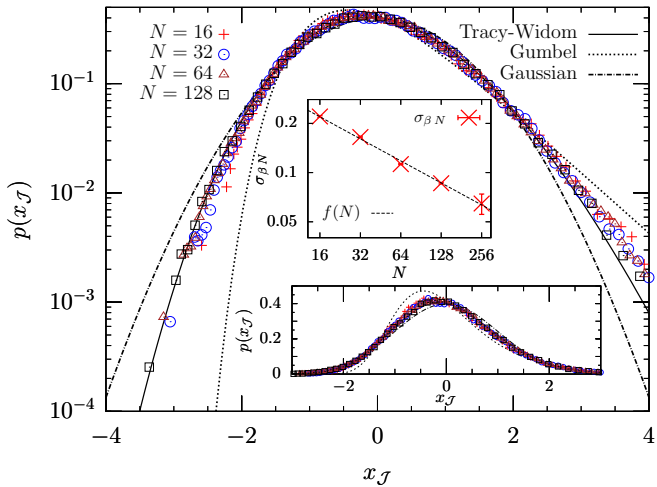}
\caption{Distribution $p(x_{\mathcal{J}})$  of the normalized critical temperature $x_{\mathcal{J}}$ for the $p$-spin model with $p=3$ and system sizes $ N = 16, 32, 64, 128$ (in red, blue, brown, black respectively), and Tracy-Widom, Gumbel and Gaussian distributions, all with zero mean  and unit variance (in black). The plot has no adjustable parameters, and it is in logarithmic scale to emphasize the shape of the distributions on the tails.  Top inset: Width $\sigma_{\beta \, N}$ of the distribution of the finite-size critical temperature as a function of the system size $N$ (in red), and fitting function $f(N) = a N^{-\phi}$ (in black), with fitting exponent $\phi = 0.45   \pm  0.04$. Bottom inset: Same plot as in the main panel in linear scale. 
\label{fig3}}
\end{figure}

The definition (\ref{eq5})  of finite-size critical temperature is depicted in Fig. \ref{fig2}. As shown in the top inset of Fig. \ref{fig3}, the variance $\sigma_{\beta \, N}^2  \equiv \overline{\beta_{c \, \mathcal{J}}^2} - \overline{\beta_{c \, \mathcal{J}}}^2$ of the critical-temperature distribution is a decreasing function of the system size $N$, in particular $\sigma_{\beta \, N} \sim N^{-\phi}$, with $\phi = 0.45   \pm  0.04$.  Indeed, $\sigma_{\beta \, N}$ represents the width of the critical region of a system with size $N$:  As the system size gets larger and larger, the width $\sigma_{\beta \, N}$ shrinks, and the  finite-size critical temperature converges to the infinite-size value $\beta_c$ (see Supplemental Material) \cite{castellana2011role,castellana2011extreme,fyodorov2012critical, billoire2011finite}. Given that that the distribution of $\beta_{c \, \mathcal{J}}$ obeys the above scaling with the system size $N$, one can expect \cite{castellana2011extreme,billoire2011finite} that the distribution of the normalized critical temperature $x_{\mathcal{J}} \equiv (\beta_{c \, \mathcal{J}} - \overline{\beta_{c \, \mathcal{J}}}) / \sigma_{\beta \, N}$, converges to a finite  limiting shape as the system size goes to infinity. This claim is supported by the numerical data shown in the main panel and bottom inset of Fig. \ref{fig3}, showing that the distribution of $x_{\mathcal{J}}$ appears to converge to a limiting distribution for large $N$. In contrast to the REM, the shape of this limiting distribution does not seem to be the Gumbel distribution, but another EVS distribution recently discovered in the theory of random matrices, the Tracy-Widom (TW) distribution \cite{tracy2002proceedings}, as shown in the main panel and bottom inset of Fig. \ref{fig3} and as suggested by statistical-hypothesis tests (see Supplemental Material). Interestingly, the TW distribution has been recently shown to play a role in describing the critical behavior of disordered systems: For example, the TW distribution  characterizes  the average number of minima in a simple model of a random-energy landscape close to its freezing transition \cite{fyodorov2012critical}, and  the finite-size fluctuations of the critical temperature in mean-field spin glasses \cite{castellana2011role,castellana2011extreme}.

\begin{figure}
\vspace{5mm}
\includegraphics[scale=0.52]{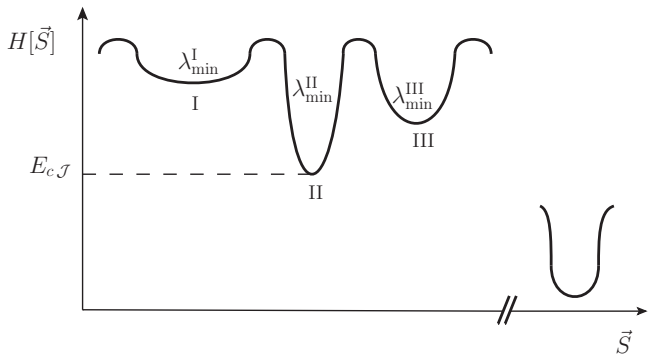}
\caption{
Schematic energy landscape of the spherical $p$-spin model: Energy $H[\vec{S}]$ of a given sample $\mathcal{J}$ as a function of the spin configuration $\vec{S}$. Three local energy minima labeled as $\textrm{I}, \textrm{II}, \textrm{III}$ (left), and one low-lying minimum (right) are depicted.  The energy of minimum $\textrm{II}$ coincides with the critical energy $E_{c \, \mathcal{J}}$ of the sample. 
The smallest eigenvalues of the   Hessian matrix $\partial^2 H / (\partial S_i \partial S_j)$ computed in each local minimum are  $\lambda_{\textrm{min}}^{\textrm{I}} , \lambda_{\textrm{min}}^{\textrm{II}},\lambda_{\textrm{min}}^{\textrm{III}}$. The geometry of the minima sets the energy threshold: Since here $\lambda_{\textrm{min}}^{\textrm{I}} < \lambda_{\textrm{min}}^{\textrm{III}} < \lambda_{\textrm{min}}^{\textrm{II}}$,  minimum $\textrm{II}$ has the largest  curvature,  thus the lowest energy amongst the three local energy minima.
\label{fig4}}
\end{figure}

We will now provide an analytical argument to give insight into the above finding that the finite-size critical temperature fluctuates according to the TW distribution. Let us consider a version of the PSM where spins $S_i$ are not binary, but continuous variables satisfying the spherical constraint $\sum_{i=1}^N S_i^2 = N$ \cite{crisanti2003complexity}. This model is known as the spherical PSM, and it has exactly the same behavior as the PSM with Ising spins above \cite{billoire2005mean}, but it is more suitable for analytical studies \cite{castellani2005spin}.  Let us now consider this PSM for any finite $p>2$ close to the transition point: The system is in the high (low) temperature phase if the average internal energy $E$  is higher (lower) than the average energy $E_c$ of the local energy minima \cite{castellani2005spin}. Now, consider a finite-size sample $\mathcal{J}$ of the PSM. If the system's internal energy $E$  is larger than the energy of the lowest local minimum, the system is in the high-temperature phase where it explores exponentially many local minima. Otherwise, the system is in the low-temperature phase where it explores the  low-lying minima \cite{castellani2005spin}. Hence, the critical energy $E_{c \, \mathcal{J}}$ of this sample is given by the energy of the lowest local minimum, as illustrated in Fig. \ref{fig4}.  Since the spin variables are continuous, we can introduce the Hessian matrix of the Hamiltonian $\partial^2 H / (\partial S_i \partial S_j)$ evaluated  in a local minimum, and its smallest eigenvalue $\lambda_{\textrm{min}}$. The fluctuations of $E_{c \, \mathcal{J}}$ can be understood by considering the geometry of the local energy minima. Indeed,  following \cite{biroli2009random} the deeper the minimum, the larger the curvature of  $H[\vec{S}]$ vs. $\vec{S}$ at the minimum, and so the larger $\lambda_{\textrm{min}}$, as illustrated in Fig. \ref{fig4}: Let us then assume \cite{biroli2009random} that the energy $E_{c \, \mathcal{J}}$ of the lowest-lying minimum is set by the smallest eigenvalue $\lambda_{\textrm{min}}$ of the Hessian in such minimum. Since for any $p > 2$ the Hessian is  a random matrix belonging to the Gaussian orthogonal ensemble (GOE) \cite{crisanti2003complexity,mehta2004random}, its smallest eigenvalue $\lambda_{\textrm{min}}$ is distributed according to the TW distribution \cite{tracy2002proceedings}.  Given that $\lambda_{\textrm{min}}$ is TW distributed, the above geometrical argument shows that the fluctuations of the critical energy $E_{c \, \mathcal{J}}$, and consequently the fluctuations of the critical temperature  $\beta_{c \, \mathcal{J}}$, are described by the TW distribution. Importantly, this argument shows that the emergence of the TW distribution shown in the numerical simulations for $p=3$ holds for any finite $p>2$. Finally, let us discuss the large-$p$ limit: In this limit, the PSM is equivalent to the REM \cite{derrida1980randomlimit}, and the distribution of the finite-size critical temperature converges to the Gumbel distribution. Indeed, for large $p$ the above local geometrical structure of the energy landscape (see Fig. \ref{fig4}) disappears, and the local minima become simply a set of IID  Gaussian random variables. Given that the energy threshold $E_{c \, \mathcal{J}}$ is the minimum of these IID Gaussian random variables, $E_{c \, \mathcal{J}}$ is distributed according to the Gumbel distribution \cite{david2003order}, and so is the critical temperature.

\textit{Conclusions} - In this Letter we have studied the finite-size fluctuations of the freezing-transition temperature of two mean-field models of structural glasses: The random energy model (REM) \cite{derrida1980randomlimit} and the $p$-spin model (PSM) \cite{gross1984simplest} with $p=3$. We find that for both the REM and the PSM, the finite-size fluctuations of the critical temperature are described by extreme-value-statistics  probability distributions. For the REM, the critical-temperature fluctuations are described by the Gumbel distribution \cite{david2003order}, while for the PSM the critical temperature is distributed according to the Tracy-Widom (TW) distribution, which has been recently discovered in the theory of random matrices \cite{tracy2002proceedings}. For the PSM, we have provided an analytical argument to understand the emergence of the TW distribution.  Recent studies  \cite{castellana2011role,castellana2011extreme} have shown that the TW distribution also emerges in spin glasses: The above analytical argument shows that the physics underlying the emergence of the TW distribution in structural glasses is completely different from spin glasses, because it involves a different mechanism related to the statistics of glassy metastable states.   Taken together, the results provided in this Letter establish an unprecedented connection between the theory of extreme values and the freezing transition of structural glasses.

As a topic of future studies, it would be interesting to study the fluctuations of the critical temperature in structural-glass models with short-range interactions, such as the Edwards-Anderson (EA) model in an external magnetic field \cite{drossel2000spin}. In this regard, previous studies have shown that EVS distributions play a role in short-range systems with quenched disorder, such as the EA model with no external field \cite{castellana2011extreme}. In fact, one could imagine short-range systems to behave as an ensemble of nearly independent subsystems, each subsystem having its own critical temperature: The finite-size critical temperature of the system as a whole is then given by the smallest of the subsystems' critical temperatures \cite{castellana2011extreme}. This raises the possibility that the crtical temperature of the system as a whole could be distributed according one of the three EVS distributions of independent and identically distributed random variables: The Gumbel, Fr\'{e}chet, or Weibull distribution, as predicted by the extreme-value theorem \cite{david2003order}.   

\paragraph*{Acknowledgments} - We would like to thank C. Broedersz and G. Schehr for useful discussions. Research supported by NSF Grants No. PHY–0957573 and CCF–0939370, by the Human Frontiers Science Program, by the Swartz Foundation, and by the W. M. Keck Foundation. Simulations were performed on computational resources supported by the Office of Information Technology's High Performance Computing Center at Princeton University.

\addcontentsline{toc}{section}{\refname}
\bibliography{bibliography}

\end{document}


\title{Supplemental Material for ``Extreme-Value Distributions and the Freezing Transition of Structural Glasses"}

\author{Michele Castellana}
\email{michelec@princeton.edu}
\affiliation{Joseph Henry Laboratories of Physics and Lewis--Sigler Institute for Integrative Genomics, Princeton University, Princeton, New Jersey 08544}

\maketitle

\section*{Convergence to the infinite-size critical temperature}

\begin{figure}[h]
\hspace{-1cm}
\includegraphics[scale=2.5]{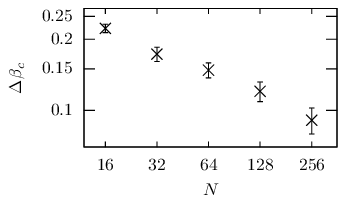}
\caption{
Deviation $\Delta \beta_c$ of the finite-size critical temperature $\beta_{c \, \mathcal{J}}$ from the infinite-size critical temperature $\beta_c$ as a function of the system size $N$. 
\label{figS1}}
\end{figure}

 In this Section we show that the finite-size critical temperature for the $p$-spin model defined in Eq. (1) converges to the infinite-size critical temperature $\beta_c =   1.535$ \cite{billoire2005mean}. To do so, we compute the quantity $\Delta \beta_c \equiv [\overline{(\beta_{c \, \mathcal{J}} - \beta_c)^2}]^{1/2}$, representing the deviation of the finite-size critical temperature $\beta_{c \, \mathcal{J}}$ from the infinite-size critical temperature $\beta_c$: The smaller $\Delta \beta_c $, the more peaked the distribution of $\beta_{c \, \mathcal{J}}$ around $\beta_c$.  Fig. \ref{figS1} shows that $\Delta \beta_c$ is a decreasing function of the system size $N$: Hence, the numerical data suggest that for large $N$ the finite-size critical temperature $\beta_{c \, \mathcal{J}}$ converges to its infinite-size value $\beta_c$. \\

\section*{Statistical-hypothesis test}

\begin{figure}
\vspace{5mm}
\centering \includegraphics[scale=1.2]{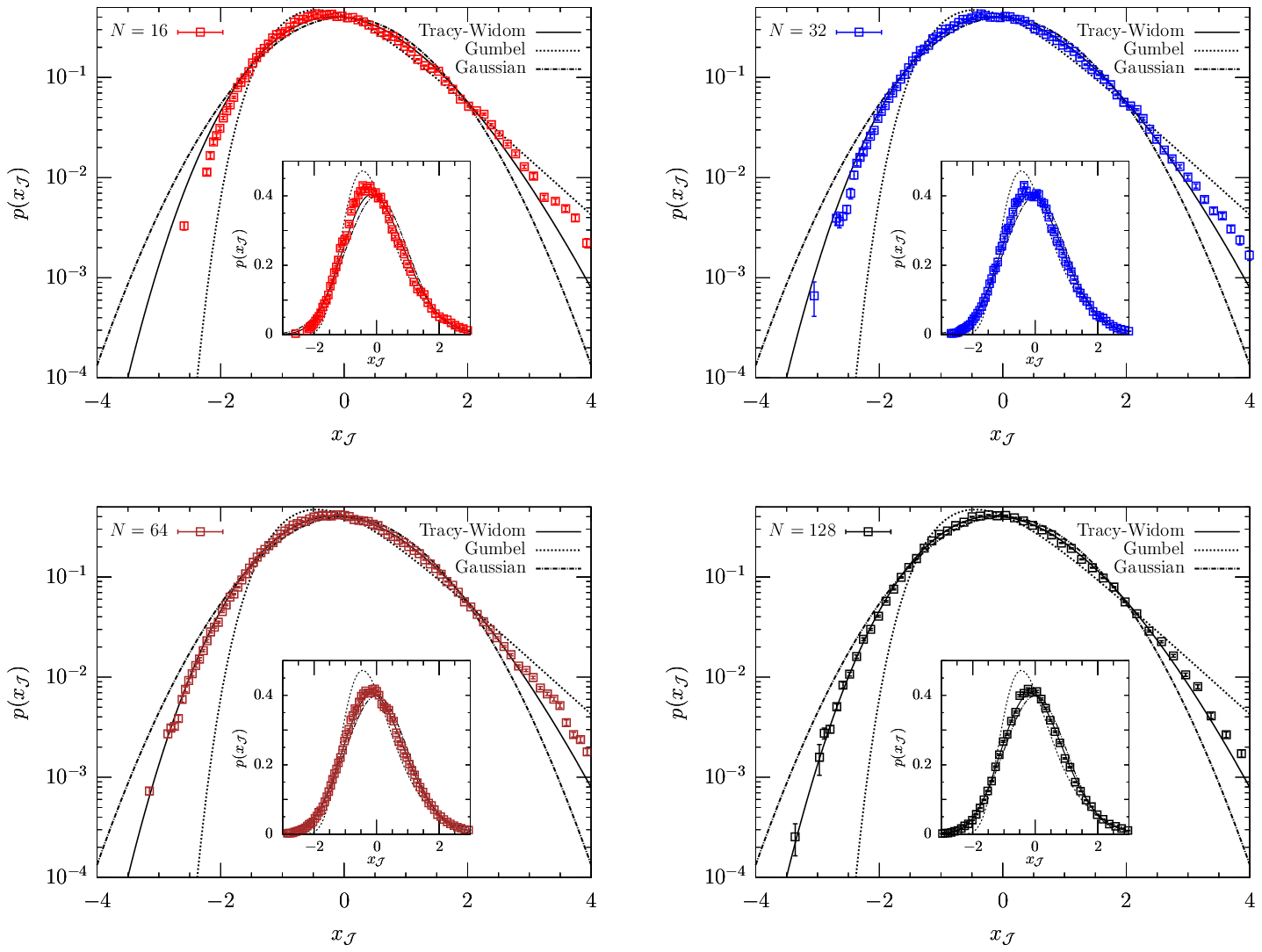}
\caption{
Distribution $p(x_{\mathcal{J}})$  of the normalized critical temperature $x_{\mathcal{J}}$ with error bars, for different system sizes $N = 16, 32, 64, 128$ (in red, blue, brown, black respectively). Each system size is represented in a different panel, together with the Tracy-Widom, Gumbel and Gaussian distribution, all with zero mean and unit variance. Insets: Same plots as in the main panels in linear scale. 
\label{figS2}}
\end{figure}

 In this Section we present a statistical test supporting the emergence of the Tracy-Widom (TW) distribution as the distribution of the finite-size critical temperature of the $p$-spin model. 
First,  in Fig. \ref{figS2} we show $p(x_{\mathcal{J}})$ for multiple system sizes $N$ with the relative error bars, which have been estimated with the bootstrap method \cite{efron1979bootstrap}: This Figure shows how the distribution $p(x_{\mathcal{J}})$  compares to the TW distribution relatively to the numerical error bars. Second, the samples of normalized finite-size critical temperatures $x_{\mathcal{J}}$ have been compared with the TW, Gumbel and Gaussian distribution with zero mean and unit variance by using the Kolmogorov-Smirnov (KS) test \cite{mood1950introduction}. The KS statistic is $D = \sup_x |F_{\mathcal{J}}(x) - F(x)|$, where $F_{\mathcal{J}}$ is the cumulative distribution function (CDF) of $x_{\mathcal{J}}$, $F$ is the CDF of the TW, Gumbel or Gaussian distribution with zero mean and unit variance. The $p$-value is $ \textrm{Pr}(x \geq \sqrt{S} D)$, where $x$ is a random variable distributed according to the Kolmogorov distribution \cite{mood1950introduction} and $S$ is the number of samples. Table \ref{tabS1} shows the $p$-value for system sizes $N = 16, 32, 64, 128$ shown in Fig. \ref{figS1}. At a significance level of $0.1$, the data do not indicate that the hypothesis that the finite-size critical temperature is distributed according to the TW distribution should be rejected for $N > 16$, while at the same significance level the hypothesis that $\beta_{c \, \mathcal{J}}$ is distributed according to the Gaussian or Gumbel distribution is rejected for all the values of $N$ shown in Table \ref{tabS1}.

\begin{table}
\begin{tabular}{|c|c|c|c|c|}
\hline
$N$ & $S$ & $p_{\textrm{TW}}$ & $p_{\textrm{Gumbel}}$ & $p_{\textrm{Gaussian}}$ \\ \hline 
16 & $1.3\times 10^5$ & $1.1 \times 10^{-5}$ & \multirow{4}{*}{$< 10^{-10}$} & \multirow{4}{*}{$< 10^{-10}$} \\
32 & $1.3\times 10^5$ & $0.26$ & &  \\
64 &$1.3\times 10^5$  & $0.25$ & & \\
128 & $6.9 \times 10^4 $ & $0.88$ & & \\
\hline
\end{tabular}
\caption{
Kolmogorov-Smirnov (KS) test for the distribution of the finite-size  critical temperature: For system sizes $N = 16, 32, 64, 128$ we show the number of samples $S$ and the $p$-value of the KS test which compares the distribution of  the normalized critical temperature with a reference probability distribution: The Tracy-Widom, Gaussian or Gumbel distribution, all with zero mean and unit variance.  Small value of the $p$-value suggest that the distribution of the normalized critical temperature is significantly different from the reference distribution. 
\label{tabS1}}
\end{table}

\bibliography{bibliography}